\documentclass[preprint,showpacs,preprintnumbers,amsmath,amssymb]{revtex4}
\usepackage{epsfig}
\usepackage{graphicx}% Include figure files
\usepackage{dcolumn}% Align table columns on decimal point
\usepackage{bm}% bold math
\usepackage{amsmath}

\def\beq{\begin{equation}}
\def\eeq#1{\label{#1}\end{equation}}
\def\eeqn{\end{equation}}
\newcommand\iden{\leavevmode\hbox{\small1\normalsize\kern-.33em1}}

\let\jnfont=\rm
\def\NPB#1,{{\jnfont Nucl.\ Phys.\ B }{\bf #1},}
\def\PLB#1,{{\jnfont Phys.\ Lett.\ B }{\bf #1},}
\def\EPJC#1,{{\jnfont Eur.\ Phys.\ Jour.\ C }{\bf #1},}
\def\PRD#1,{{\jnfont Phys.\ Rev.\ D }{\bf #1},}
\def\PRL#1,{{\jnfont Phys.\ Rev.\ Lett.\ }{\bf #1},}
\def\MPLA#1,{{\jnfont Mod.\ Phys.\ Lett.\ A }{\bf #1},}
\def\JPG#1,{{\jnfont J.\ Phys.\ G }{\bf #1},}
\def\CTP#1,{{\jnfont Commun.\ Theor.\ Phys.\ }{\bf #1},}
\def\JHEP#1,{{\jnfont JHEP \ }{\bf #1},}
\def\NPPS#1,{{\jnfont Nucl.\ Phys.\ Proc.\ Suppl.\ }{\bf #1},}
\def\CPC#1,{{\jnfont Computl.\ Phys.\ Commun.\ }{\bf #1},}

\begin{document}

\preprint{\parbox{1.2in}{\noindent arXiv:0712.???? }}

\title{\ \\[10mm] Top-Quark FCNC Decay $t\to cgg$ \\ in Topcolor-assisted Technicolor Model}

\author{\ \\[2mm] Huan-Jun Zhang \\ ~}

\affiliation{Department of Physics, Henan Normal University,
Xinxiang, Henan 453007, China \vspace*{1.5cm} }

\begin{abstract}
The topcolor-assisted technicolor (TC2) model predicts several pseudo-scalars called  
top-pions and at loop level they can induce the FCNC top quark decay $t\to cgg$ 
which is extremely suppressed in the Standard Model (SM). 
We find that in the allowed parameter space the TC2 model can greatly enhance 
such a FCNC decay and push the branching ratio up to $10^{-3}$, which
is much larger than the predictions in the SM ($10^{-9}$) and in the
minimal supersymmetric model ($10^{-4}$). 
We also compare the result with the two-body FCNC decay $t\to cg$ and
find that the braching ratio of $t\to cgg$ is slightly larger than $t\to cg$.
Such enhanced FCNC top quark decays may serve as a good probe of TC2 model
at the future top quark factory. 
\end{abstract}

\maketitle

{\em Introduction:}~
The upcoming Large Hadron Collider (LHC) will put various new physics ideas 
to the sword. While this machine has enough energy to produce TeV-scale new 
particles and thus can directly probe TeV-scale new physics, one should 
also pay sufficient attention to the indirect probe through revealing quantum 
effects of new physics in some sensitive processes.        
As the heaviest fermion in the Standard Model (SM), the top quark is speculated 
to be a sensitive probe of new physics \cite{top-review}. So far the top quark 
properties are not precisely measured due to the small statistics of the experiments 
at the Fermilab Tevatron collider. The LHC and the proposed International Linear Collider 
(ILC) will copiously produce top quarks and allow to scrutinize the top quark nature.

One of the properties of the top quark in the SM is its
extremely small flavor-changing neutral-current (FCNC)
interactions  \cite{tcvh-sm} due to the Glashow-Iliopoulos-Maiani (GIM)
mechanism. This will make the observation of any
FCNC top quark prosess a robust evidence for new physics
beyond the SM.
So far numerous studies \cite{top-fcnc-review} have shown that the FCNC top quark interactions 
can be significantly 
enhanced in some new physics models like the minimal supersymmetric model (MSSM) 
\cite{tcv-pptc-mssm,tcv-mssm,pptc-mssm} and the TC2 model \cite{tcv-TC2,pptc-TC2}. 
It was found that for these FCNC
top quark processes the TC2 model usually allows for much larger production or decay rates 
than the supersymmetric model. Through the measurements of the FCNC top quark 
processes at the LHC or ILC, the effects of these new physics models will be revealed. 

Although so far in the literature there are several works devoted to
the TC2 contributions to the FCNC top quark decays, the TC2 prediction
for the three-body decay $t\to cgg$ has not been studied yet.
As shown in \cite{tcvh-sm,tcv-pptc-mssm}, in both the SM and MSSM this decay is 
found to have a larger branching ratio than the two-body decay $t\to cg$ 
and thus may be the most hopeful FCNC top decay channel to discover 
at the LHC or ILC.  In this work we focus on the TC2 contribution to this decay
and compare its branching ratio with $t\to cg$. 

This work is organized as follows. 
We will first discuss the TC2 model and then perform the calculations. 
Since the calculations involve many loops and are somewhat tedious, we will not 
present the details and will instead give the analytical results in an appendix.
We will present some numerical examples with comparison to the results in the
SM and MSSM, and finally give our conclusion.        
\vspace*{0.5cm}
  
{\em Calculations:}~
Before our calculations we recapitulate the basics of the TC2
model. As is well known, the fancy idea of technicolor aims to 
dynamically break the electroweak symmetry, but it encounters enormous 
difficulty in generating fermion masses, especially the heavy top quark 
mass. The TC2 model \cite{TC2-model} combines technicolor with
top-color, with the former being responsible for
electroweak symmetry breaking and the latter for generating large
top-quark mass. This model so far survives current experiments
and awaits being tested at the LHC. 

A crucial aspect of TC2 phenomenology will be related to the light 
pseudo-Goldstone bosons called the top-pions ($\pi_t^0$ and $\pi_t^\pm$),
which are predicted in TC2 model at the weak scale \cite{TC2-model} and
have flavor-changing couplings with the top quark 
\begin{eqnarray}
&\frac{(1-\epsilon )m_t}{\sqrt{2}F_{t}}\frac{\sqrt{v^2-F_{t}^2}}{v}&
  \left( iK_{UL}^{tt}K_{UR}^{tt}\bar{t}_Lt_R\pi_t^0 
  +iK_{UL}^{tt}K_{UR}^{tc}\bar{t}_Lc_R\pi_t^0 \right.\nonumber\\ 
&& +\sqrt{2}K_{UR}^{tt}K_{DL}^{bb}\bar{t}_Rb_L\pi_t^-
   +\sqrt{2}K_{UR}^{tc}K_{DL}^{bb}\bar{c}_Rb_L\pi_t^- \nonumber\\  
&& \left.+ K_{UL}^{tt}K_{UR}^{tt}\bar{t}_Lt_Rh_t^0
 + K_{UL}^{tt}K_{UR}^{tc}\bar{t}_Lc_Rh_t^0 +h.c.\right), \label{eq1}
\end{eqnarray}
where the factor $\sqrt{v^2-F_t^2}/v$ ( $v \simeq 174$ GeV )
reflects the effect of the mixing between the top pions and the
would-be Goldstone bosons. The parameter $\epsilon$ parameterizes 
the portion of the extended-technicolor contribution to the top quark 
mass. $K_{UL}$,$K_{DL}$ and $K_{UR}$ are the rotation matrices that 
transform respectively the weak eigenstates of left-handed up-type, 
down-type, and right-handed up-type quarks to their mass eigenstates, 
whose values can be parameterized as \cite{pptc-TC2}
\begin{equation}
K_{UL}^{tt}  \simeq 1, \hspace{5mm}
K_{UR}^{tt}\simeq \frac{m_t^\prime}{m_t} = 1-\epsilon,\hspace{5mm}
K_{UR}^{tc}\leq \sqrt{1-(K_{UR}^{tt})^2}
=\sqrt{2\epsilon-\epsilon^{2}}, \label{eq2}
\end{equation}
with $m_t^\prime$ denoting the top-color contribution to the top
quark mass. In our calculations we assume $K_{UR}^{tc}=\sqrt{1-(K_{UR}^{tt})^2}$.
 The TC2 model also predicts a CP-even scalar called
top Higgs ($h_t^0$) whose couplings to top quark are similar to
the neutral top pion \cite{pptc-TC2}. 
%%%%%%%%%%%%%%%%%%%%%%%%%%%%%%%%%%%%%%%%%%%%%%%%%%%%%%%%%%%%%%%%%%%%%%%%
 \begin{figure}[tb]
 \begin{center}
 \hspace*{-0.5cm}\scalebox{0.8}{\epsfig{file=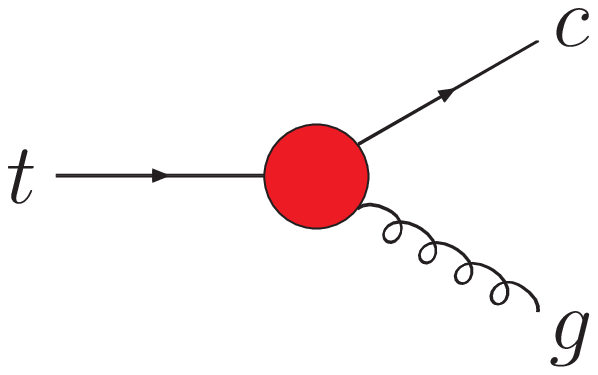,width=4cm}}
 \raisebox{5ex}[0pt]{=}\ \ \ \
 \scalebox{0.8}{\epsfig{file=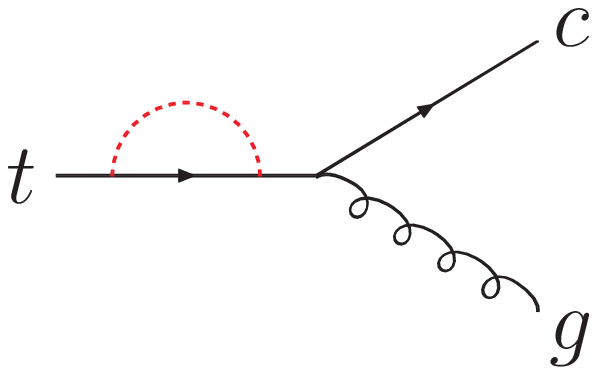,width=4cm}}\ \ \
 \raisebox{5ex}[0pt]{+}\ \ \
 \scalebox{0.8}{\epsfig{file=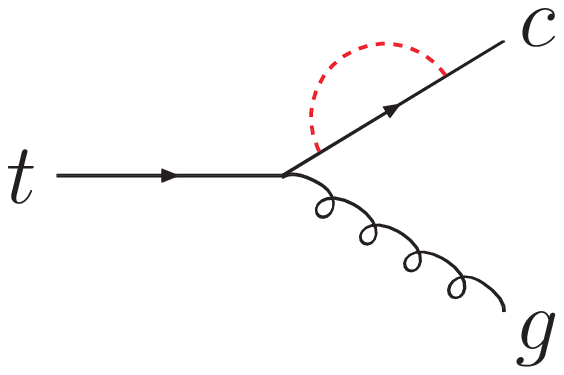,width=4cm}}\ \ \
 \raisebox{5ex}[0pt]{+}\ \ \
 \scalebox{0.8}{\epsfig{file=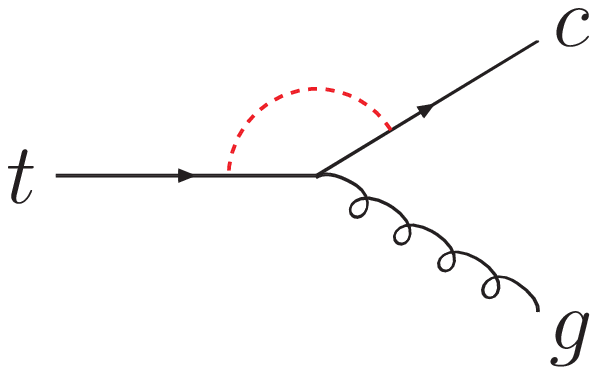,width=4cm}}
 \end{center}
\vspace{-0.5cm} \caption{Feynman diagrams for the effective vertex
$t\bar{c}g$ at one-loop level in TC2 model. The boson in each loop 
denotes a neutral top-pion, a top-Higgs or a charged top-pion, 
while the fermion in each loop is correspondingly a top or bottom quark.}
\end{figure}
%%%%%%%%%%%%%%%%%%%%%%%%%%%%%%%%%%%%%%%%%%%%%%%%%%%%%%%%%%%%%%%%%%%%%%%%%%

These flavor-changing couplings will induce the FCNC coupling 
$t\bar c g$, as shown in Fig.~1.  We follow the idea of the 'effective vetrtex' 
in \cite{tcv-pptc-mssm} and  
define an effective $t\bar{c}g$ vertex to simplify our calculations
\begin{eqnarray} \label{eff-vertex}
\Gamma_\mu^{eff}(p_t,p_c) = \Gamma_\mu^{t\bar{c}g}(p_t,p_c) 
 +\Gamma_\mu^{c\bar{c}g} \frac{i({p\!\!\slash}_t+m_c)}{p_t^2-m_c^2} i\Sigma(p_t)
 +i\Sigma(p_c) \frac{i({p\!\!\slash}_c+m_t)}{p_c^2-m_t^2} \Gamma_\mu^{t\bar{t}g},
\end{eqnarray}
where $\Gamma_\mu^{q\bar{q}g}$ ($q=c,t$) is the usual QCD vertex, 
and $\Gamma_\mu^{t\bar{c}g}$, $\Sigma(p_t)$ and  $\Sigma(p_c)$ are
respectively the contributions from vertex and self-energy loops shown
in Fig.~1, whose expressions are given in the Appendix.
Such an effective veretx can be re-shaped in the form 
\begin{eqnarray} \label{vertex-2}
F_1(k^2) T^a (k^2\gamma_\mu-k_\mu k\!\!\!\slash) 
  +m_tF_2(k^2)T^a i\sigma_{\mu\nu}k^\nu,
\end{eqnarray}
where $k$ denotes the momentum of the gluon, $T^a$ ($a=1,\cdots,8$) 
are the $SU_c(3)$ generators, 
and $F_{1,2}(k^2)$ are form factors arising from loop calculations. 
Note that for the two-body decay $t\to cg$, $F_1$ does not
contribute since the gluon momentum $k$ satisfies $k^2=0$ and
$k\cdot\epsilon=0$ with $\epsilon$ being the gluon polarization 
vector (different from $\epsilon$ in Eqs.(\ref{eq1}) and (2) !). 
%%%%%%%%%%%%%%%%%%%%%%%%%%%%%%%%%%%%%%%%%%%%%%%%%%%%%%%%%%%%%%%%%%%%%%%%
 \begin{figure}[tb]
 \begin{center}
 \scalebox{1.0}{\epsfig{file=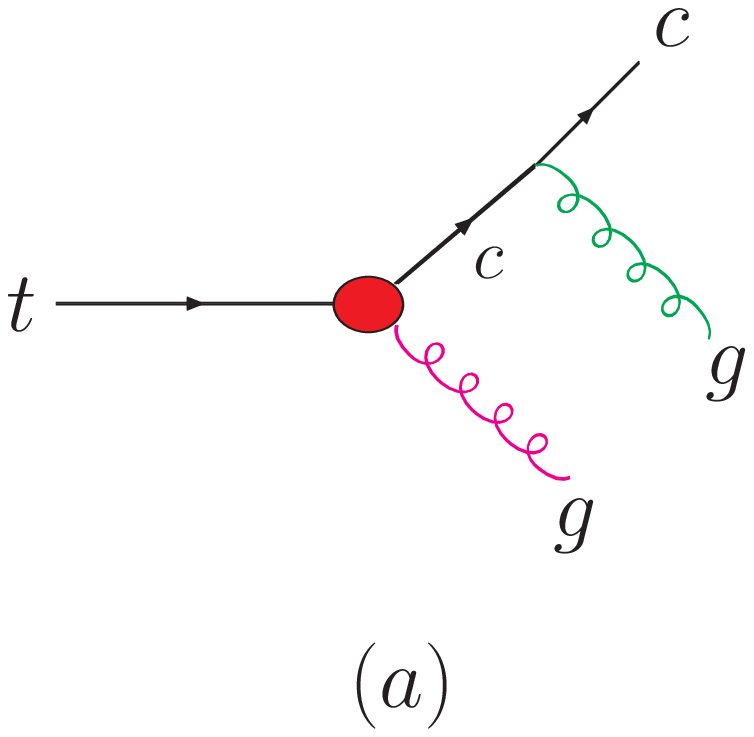,width=4cm}}\ \ \ \ \ \ \ \ \ \
 \scalebox{1.0}{\epsfig{file=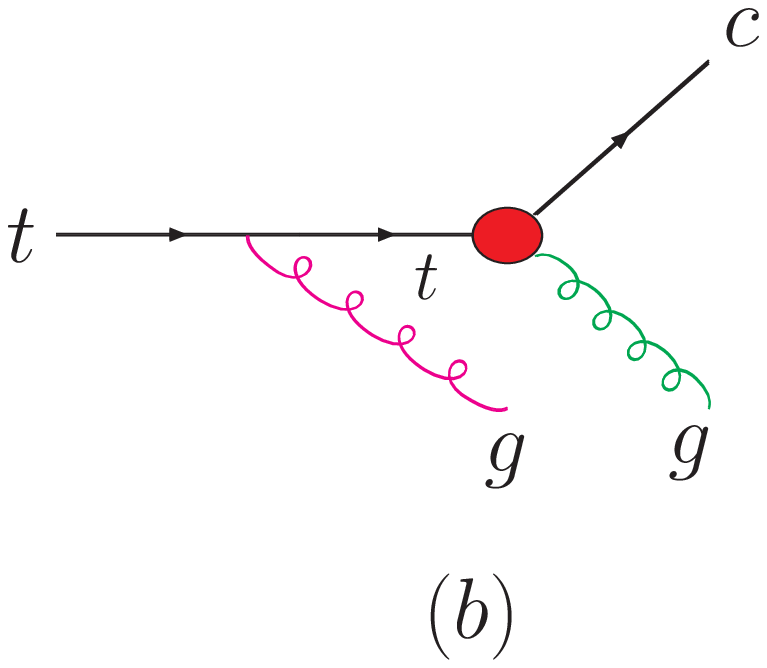,width=4cm}}
 \scalebox{1.0}{\epsfig{file=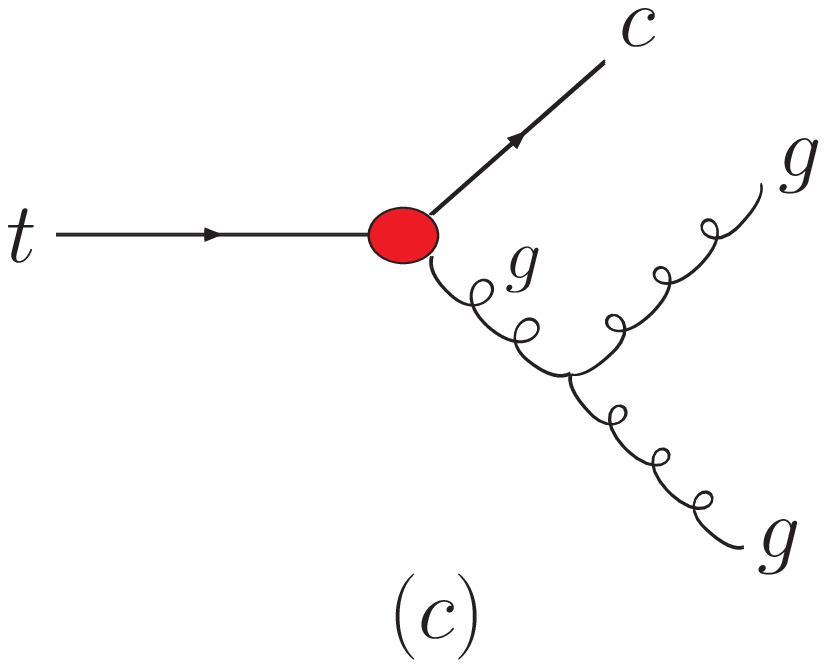,width=4cm}}

\vspace*{-1cm} \hspace*{-2cm}
 \scalebox{1.1}{\epsfig{file=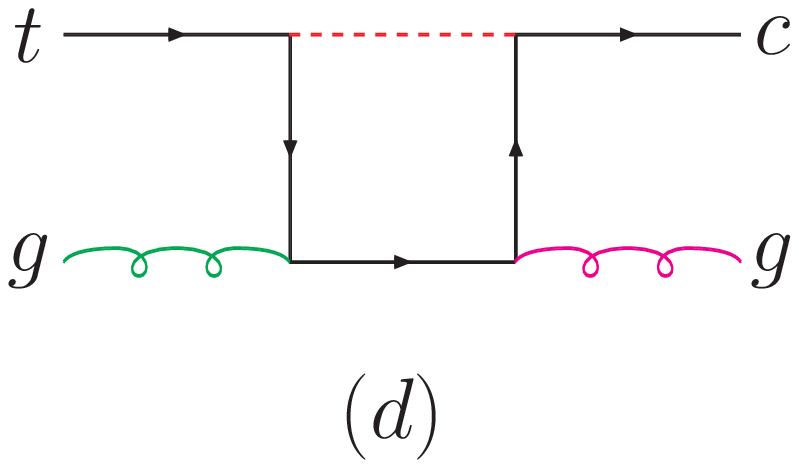,width=4cm}}\ \ \ \ \ \ \ \ \ \
 \scalebox{1.1}{\epsfig{file=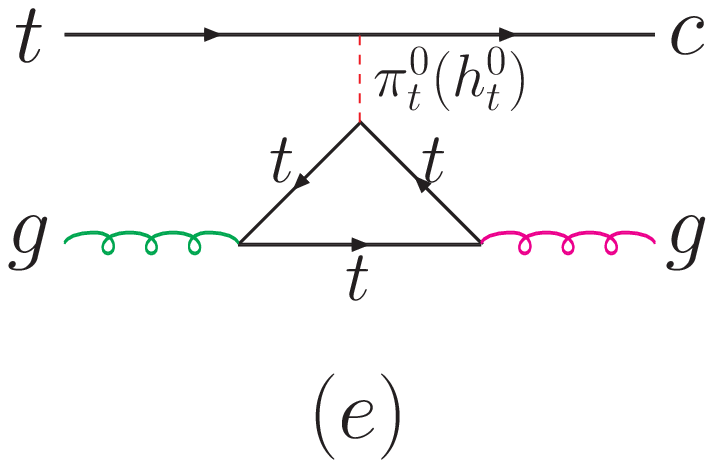,width=4cm}}
 \end{center}
\vspace{-0.5cm} \caption{Feynman diagrams for $t\rightarrow cgg$
at one-loop level in TC2 model. The effective vertex $t\bar{c}g$
in (a-c) is defined in Fig.~1. The boson in the loop of the box 
diagram (d) can be a neutral top-pion, top-Higgs or a charged top-pion, 
while the fermion in the loop can be a top or bottom quark correspondingly.
The two gluons in (a,b,d,e) can be exchanged.}
\end{figure}
%%%%%%%%%%%%%%%%%%%%%%%%%%%%%%%%%%%%%%%%%%%%%%%%%%%%%%%%%%%%%%%%%%%%%%%%%%

With the effective vertex defined above, the Feynman diagrams for
the decay $t\to cgg$ are shown in Fig.~2, where the diagrams (a-c)
involve the effective vertex. The amplitudes of the diagrams (a-c)
are obvious with the defined effective vertex. In addition, we need
to calculate the box diagrams shown in Fig.2(d) and the triangle top-quark 
loop diagram shown in Fig.2(e). The calculations are straightforward and 
the results are given in the Appendix.
\vspace*{0.5cm}
 
{\em Numerical results:}~ Now we are ready to give some numerical results.
First, we take a look at the involved parameters.  
The parameters in our calculations are the masses of the top-pions and top-Higgs,
$K_{UR}^{TC}$, and the top-pion decay constant $F_t$. 
In our calculations we take $m_t=170.9$ GeV \cite{top-mass} and $F_t = 50$ GeV.
About the top-pion and top-Higgs  masses, in our analysis we assume
\begin{equation}
m_{\pi_t^0} = m_{\pi^{\pm}_t} = m_{h_t^0} \equiv M_{TC}
\end{equation} 
In our figures of numerical results we will show a bound of about 250 GeV 
on top-pion mass, which is from $R_b$ constraint on the charged top-pion \cite{kuang}.
Note that such a bound is not so robust since in TC2 model the sizable corrections to 
$R_b$ can also come from the extended technicolor gauge bosons. 
%%%%%%%%%%%%%%%%%%%%%%%%%%%%%%%%%%%%%%%%%%%%%%%%%%%%%%%%%%%%%%%%%%%%%%%%%%%%
\begin{figure}[tb]
\begin{center}
\scalebox{0.9}{\epsfig{file=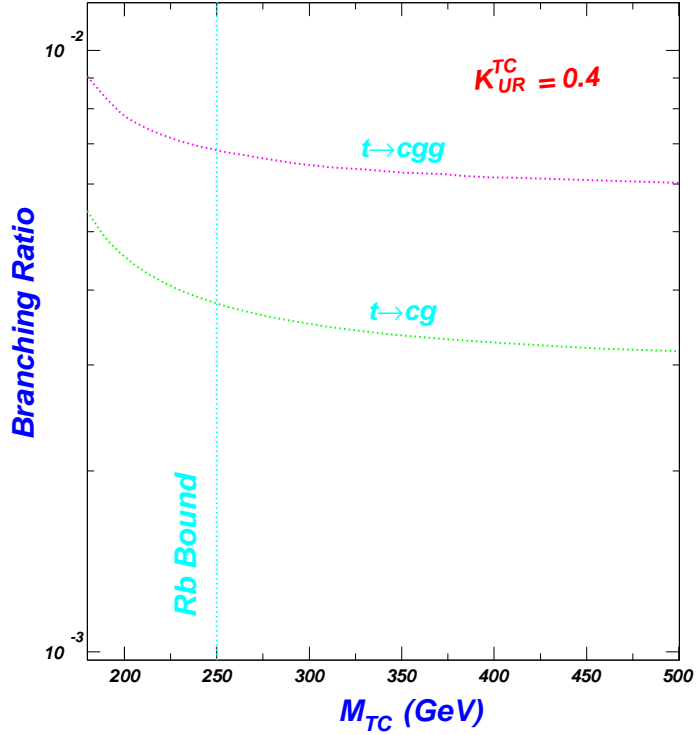,width=10cm}}
\end{center}
\vspace*{-1.0cm} \caption{Branching ratios of $t\rightarrow cgg$
and $t\rightarrow cg$ versus $M_{TC}$ in TC2 model.}
\end{figure}
%%%%%%%%%%%%%%%%%%%%%%%%%%%%%%%%%%%%%%%%%%%%%%%%%%%%%%%%%%%%%%%%%%%%%%%%%%%%
%%%%%%%%%%%%%%%%%%%%%%%%%%%%%%%%%%%%%%%%%%%%%%%%%%%%%%%%%%%%%%%%%%%%%%%%%%%%
\begin{figure}[tb]
 \scalebox{0.9}{\epsfig{file=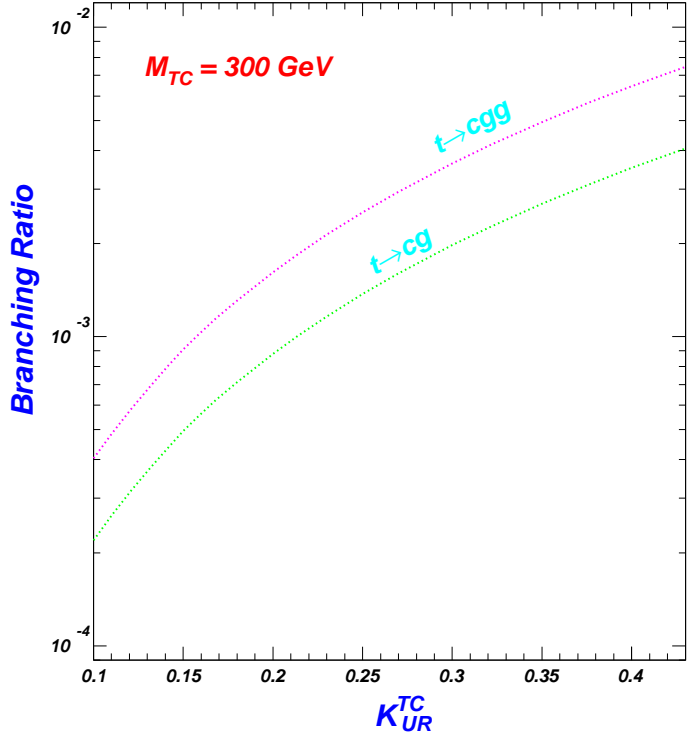,width=10cm}}
\vspace*{-0.6cm} 
\caption{Branching ratios of $t\rightarrow cgg$
and $t\rightarrow cg$ versus $K_{UR}^{TC}$ in TC2 model.}
\end{figure}
%%%%%%%%%%%%%%%%%%%%%%%%%%%%%%%%%%%%%%%%%%%%%%%%%%%%%%%%%%%%%%%%%%%%%%%%%%%%%%

In our numerical results we give the branching ratio with
the top width taken to be $\Gamma_t = 1.55$ GeV \cite{pptc-mssm}. 
To make our predictions more realistic, we
apply some kinematic cuts as in \cite{tcv-pptc-mssm}, e.g., 
we require the energy of each decay
product be larger than 15 GeV in the top quark rest frame.

In Fig.3 we show the branching ratios of $t\to cgg$ and 
$t\to cg$ versus $M_{TC}$. We see that the contributions of top-pions can 
significantly enhance such rare decays and in the allowed parameter space 
the branching ratio can be up to $10^{-3}$, which
is much larger than the predictions in the SM ($10^{-9}$) \cite{tcvh-sm} 
and in the MSSM ($10^{-4}$) \cite{tcv-pptc-mssm}. 

As shown in Fig.3, the branching ratio of  $t\to cgg$ is
larger than $t\to cg$, which is also observed in the
SM \cite{tcvh-sm} and the MSSM \cite{tcv-pptc-mssm}. 
As discussed in \cite{tcvh-sm,tcv-pptc-mssm}, 
the reason for this behavior is that the form factor $F_1$ in Eq.(\ref{vertex-2}), which 
makes important contribution to $t\to cgg$,  does not contribute to $t\to cg$. 

Note that the TC2 contributions are sensitive to the parameter $K_{UR}^{TC}$
which is fixed to 0.4 in Fig.3.   
In Fig.4 we show the dependence on $K_{UR}^{TC}$ for fixed top-pion mass. 
We see that the branching ratios increase drastically as $K_{UR}^{TC}$ 
gets large.

Finally, in Table 1 we summarize the TC2 predictions for the FCNC top 
quark decays with comparison to the predictions in the SM and MSSM.
The TC2 predictions are taken from Fig.3 for $M_{TC}=300$ GeV and
$K_{UR}^{TC}=0.4$. We see that for each decay mode the TC2 model 
allows a much larger branching ratio than the other two models.

%%%%%%%%%%%%%%%%%%%%%%%%%%%%%%%%%%%%%%%%%%%%%%%%%%%%%%%%%%%%%%%%%%%%%%%%%%%%%
%\begin{center}
\null \noindent{\small Table 1: Predictions for the branching
ratios of the FCNC top quark decays. The TC2 predictions are 
taken from Fig.3 for $M_{TC} = 300$ GeV and $K_{UR}^{TC}=0.4$.
The MSSM predictions are the maximal values in the allowed 
parameter space.}  
%\end{center}
\vspace*{-0.3cm}

\begin{center}
\doublerulesep 1.5pt \tabcolsep 0.2in
\begin{tabular}{||c|c|c|c||}\hline\hline
 & SM & MSSM & TC2\\ \hline
$Br(t\rightarrow cZ)$ & $O(10^{-13})$\cite{tcvh-sm} & $O(10^{-7})$\cite{tcv-pptc-mssm} & $O(10^{-4})$\cite{tcv-TC2}\\
\hline
$Br(t\rightarrow c\gamma)$ & $O(10^{-13})$\cite{tcvh-sm} & $O(10^{-7})$\cite{tcv-pptc-mssm} & $O(10^{-6})$\cite{tcv-TC2}\\
\hline
$Br(t\rightarrow cg)$ & $O(10^{-11})$\cite{tcvh-sm} &$O(10^{-5})$\cite{tcv-pptc-mssm} & $O(10^{-3})$\\
\hline
$Br(t\rightarrow cgg)$ & $O(10^{-9})$\cite{tcvh-sm} &
$O(10^{-4})$\cite{tcv-pptc-mssm} & $O(10^{-3})$
\\ \hline\hline
\end{tabular}
\end{center}
%%%%%%%%%%%%%%%%%%%%%%%%%%%%%%%%%%%%%%%%%%%%%%%%%%%%%%%%%%%%%%%%%%%%%%%%%%%%%%%%%%%%

In conclusion, we evaluated the TC2 contributions to the top quark FCNC decay 
$t\to cgg$ with comparison to $t\to cg$. We found that the branching ratios 
for these two decays can be enhanced to the level of $10^{-3}$, which
is much larger than the predictions in both the SM and MSSM. 
As in the SM and MSSM, the decay $t\to cgg$ was found to have a larger  
branching ratio than two-body decay $t\to cg$. 
The future precision study of the top quark properties at the LHC or ILC, 
especially the measurement of various rare decay modes, will shed some
light on the TC2 model. 
\vspace*{0.5cm}

%\section*{Acknowledgment}
We thank Junjie Cao, Jin Min Yang and Xuelei Wang for discussions, and   
Wenyu Wang and Lei Wang for help with the fortran codes. 

\appendix

\section{Expressions of loop results}

The expressions of $\Gamma_\mu^{t\bar{c}g}$, $\Sigma(p_t)$ and  $\Sigma(p_c)$ 
in the effective vertex of Eq.(\ref{eff-vertex}) are given by 
\begin{eqnarray}
\Gamma_\mu^{t\bar{c}g}&=& \frac{1}{2}ag_s^2P_L 
   \left[\gamma^\rho\gamma^\mu\gamma^\lambda(C_{\rho\lambda}^1
   +C_{\rho\lambda}^2 + 2C_{\rho\lambda}^3) +\gamma^\rho\gamma^\mu
     k\!\! \!\slash(C_\rho^1 + C_\rho^2 + 2C_\rho^3)\right. \nonumber\\
&& +m_t\gamma^\rho\gamma^\mu(-C_\rho^1+C_\rho^2)
   +m_t\gamma^\mu\gamma^\rho(-C_\rho^1+C_\rho^2)
   +m_t\gamma^\mu k\!\!\!\slash(-C_0^1+C_0^2)\nonumber\\
&& \left.+m_t^2\gamma^\mu(-C_0^1+C_0^2)\right]\\
 i\Sigma(p_t) &=&
 \frac{1}{2}aP_L[{p\!\!\slash}_t(B_1^1 +B_1^2 + 2B_1^3)+m_t(-B_0^1+B_0^2)]\\
i\Sigma(p_c)&=&\frac{1}{2}aP_L[{p\!\!\slash}_c (B_1^4 +B_1^5 + 2B_1^6)+m_t(-B_0^4 +B_0^5)]
\end{eqnarray}
with $p_t$, $p_c$ and $k$ denoting respectively the momenta of the top, charm quark and
gluon, $a=\frac{m_t^2}{F_t^2}\frac{\upsilon^2-F_t^2}{\upsilon^2}K_{UR}^{TC}$ and the loop
functions' dependence given by 
\begin{eqnarray}
&& B^1=B(p_t^2, m_t^2 , m_{\pi_t^0}^2),~B^2=B(p_t^2, m_t^2, m_{h_t^0}^2),
   ~B^3=B(p_t^2, m_b^2, m_{\pi_t^+}^2),\\
&& B^4=B(p_c^2, m_t^2, m_{\pi_t^0}^2),~B^5=B(p_c^2, m_t^2, m_{h_t^0}^2),
   ~B^6=B(p_c^2, m_b^2, m_{\pi_t^+}^2),\\
&&C^1=C(k, -p_t, m_t^2, m_t^2,m_{\pi_t^0}^2),
  ~C^2=C(k, -p_t, m_t^2, m_t^2,m_{h_t^0}^2), \\
&&  C^3=C(k, -p_t, m_b^2,m_b^2, m_{\pi_t^+}^2)
\end{eqnarray}
The amplitudes of the box diagrams in Fig.2(d) is given by 
\begin{eqnarray}
M_1 &=& -\frac{1}{2}ag_s^2T_1\frac{i}{16\pi^2}\bar{u}(p_c)P_L\big\{
   -\gamma^\rho\gamma^\nu\gamma^\lambda\gamma^\mu\gamma^\sigma
   D_{\rho\lambda\sigma}^1-\gamma^\rho\gamma^\nu\gamma^\lambda\gamma^\mu
   {p\!\!\slash}_1 D_{\rho\lambda}^1-\gamma^\rho\gamma^\nu\gamma^
   \lambda\gamma^\mu {p\!\!\slash}_2 D_{\rho\lambda}^1\nonumber\\
&& +m_t\gamma^\rho\gamma^\nu\gamma^\lambda\gamma^\mu
   D_{\rho\lambda}^1 -\gamma^\rho\gamma^\nu{p\!\!\slash}_2
   \gamma^\mu\gamma^\lambda D_{\rho\lambda}^1 +
   m_t\gamma^\rho\gamma^\nu\gamma^ \mu\gamma^\lambda
   D_{\rho\lambda}^1 \nonumber\\
&& -\gamma^\rho\gamma^\nu{p\!\!\slash}_2\gamma^\mu {p\!\!\slash}_2
   D_{\rho}^1+m_t \gamma^\rho\gamma^\nu{p\!\!\slash}_2\gamma^\mu
   D_\rho^1-\gamma^\rho\gamma^\nu{p\!\!\slash}_2\gamma^\mu
   {p\!\!\slash}_1 D_{\rho}^1 \nonumber\\
&& +m_t\gamma^\rho\gamma^\nu\gamma^\mu {p\!\!\slash}_1 D_\rho^1
   +m_t \gamma^\rho\gamma^\nu\gamma^\mu{p\!\!\slash}_2 D_{\rho}^1
   +m_t \gamma^\nu{p\!\!\slash}_2\gamma^\mu{p\!\!\slash}_2 D_0^1\nonumber\\
&& +m_t\gamma^\nu\gamma^\rho\gamma^\mu\gamma^\lambda D_{\rho\lambda}^1
   +m_t \gamma^\nu\gamma^\rho\gamma^\mu{p\!\!\slash}_1 D_\rho^1 
   +m_t\gamma^\nu\gamma^\rho\gamma^\mu{p\!\!\slash}_2D_\rho^1\nonumber\\
&& -m_t^2 \gamma^\nu\gamma^\rho\gamma^\mu D_\rho^1
   +m_t \gamma^\nu {p\!\!\slash}_2\gamma^\mu \gamma^\rho
   D_\rho^1~+ m_t \gamma^\nu {p\!\!\slash}_2\gamma^
   \mu{p\!\!\slash}_1D_0^1\nonumber\\
&&-m_t^2\gamma^\rho\gamma^\nu\gamma^\mu D_\rho^1 
  -m_t^2 \gamma^\nu{p\!\!\slash}_2\gamma^\mu D_0^1  
  -m_t^2\gamma^\nu\gamma^\mu\gamma^\rho D_\rho^1 \nonumber\\
&& -m_t^2\gamma^\nu\gamma^\mu{p\!\!\slash}_1 D_0^1
   -m_t^2\gamma^\nu\gamma^\mu{p\!\!\slash}_2 D_0^1
   +m_t^3\gamma^\nu\gamma^\mu D_0^1\big\}u(p_t)\epsilon_\nu^*(p_2)\epsilon_\mu^*(p_1)\\
%\end{eqnarray}
%\begin{eqnarray}
 M_2 &=& \frac{1}{2}ag_s^2T_1\frac{i}{16\pi^2}\bar{u}(p_c)P_L\big\{
   \gamma^\rho\gamma^\nu\gamma^\lambda\gamma^\mu\gamma^\sigma D_{\rho\lambda\sigma}^2
  +\gamma^\rho\gamma^\nu\gamma^\lambda\gamma^\mu {p\!\!\slash}_2
  D_{\rho\lambda}^2+\gamma^\rho\gamma^\nu\gamma^ \lambda\gamma^\mu
  {p\!\!\slash}_2 D_{\rho\lambda}^2\nonumber\\
&&+m_t\gamma^\rho\gamma^\nu\gamma^\lambda\gamma^\mu
  D_{\rho\lambda}^2+\gamma^\rho\gamma^\nu{p\!\!\slash}_2
  \gamma^\mu\gamma^\lambda D_{\rho\lambda}^2 
  +m_t\gamma^\rho\gamma^\nu\gamma^ \mu\gamma^\lambda D_{\rho\lambda}^2 \nonumber\\
&&+\gamma^\rho\gamma^\nu{p\!\!\slash}_2\gamma^\mu {p\!\!\slash}_2
  D_{\rho}^2+m_t \gamma^\rho\gamma^\nu{p\!\!\slash}_2\gamma^\mu D_\rho^2
  +\gamma^\rho\gamma^\nu{p\!\!\slash}_2\gamma^\mu{p\!\!\slash}_1 D_{\rho}^2 \nonumber\\
&&+m_t\gamma^\rho\gamma^\nu\gamma^\mu {p\!\!\slash}_1 D_\rho^2
  +m_t \gamma^\rho\gamma^\nu\gamma^\mu{p\!\!\slash}_2 D_{\rho}^2
  +m_t \gamma^\nu{p\!\!\slash}_2\gamma^\mu{p\!\!\slash}_2 D_0^2\nonumber\\
&& +m_t\gamma^\nu\gamma^\rho\gamma^\mu\gamma^\lambda D_{\rho\lambda}^2
   +m_t\gamma^\nu\gamma^\rho\gamma^\mu{p\!\!\slash}_1 D_\rho^2 
   +m_t\gamma^\nu\gamma^\rho\gamma^\mu{p\!\!\slash}_2D_\rho^2\nonumber\\
&& +m_t^2 \gamma^\nu\gamma^\rho\gamma^\mu D_\rho^2
   +m_t \gamma^\nu {p\!\!\slash}_2\gamma^\mu \gamma^\rho D_\rho^2
   +m_t \gamma^\nu {p\!\!\slash}_2\gamma^\mu{p\!\!\slash}_1D_0^2\nonumber\\
&& +m_t^2\gamma^\rho\gamma^\nu\gamma^\mu D_\rho^2
   +m_t^2 \gamma^\nu {p\!\!\slash}_2\gamma^\mu D_0^2
   +m_t^2\gamma^\nu\gamma^\mu\gamma^\rho D_\rho^2 \nonumber\\
&& +m_t^2\gamma^\nu\gamma^\mu{p\!\!\slash}_1 D_0^2
   +m_t^2\gamma^\nu\gamma^\mu{p\!\!\slash}_2 D_0^2
   +m_t^3\gamma^\nu\gamma^\mu D_0^2\big\}u(p_t)\epsilon_\nu^*(p_2)\epsilon_\mu^*(p_1)\\
M_3 &=&a g_s^2T_1\frac{i}{16\pi^2}\bar{u}(p_c)P_L\big\{
  \gamma^\rho\gamma^\nu\gamma^\lambda\gamma^\mu\gamma^\sigma D_{\rho\lambda\sigma}^3
  +\gamma^\rho\gamma^\nu\gamma^\lambda\gamma^\mu {p\!\!\slash}_2 D_{\rho\lambda}^3
  +\gamma^\rho\gamma^\nu\gamma^ \lambda\gamma^\mu
  {p\!\!\slash}_2 D_{\rho\lambda}^3\nonumber\\
&&+m_t\gamma^\rho\gamma^\nu\gamma^\lambda\gamma^\mu D_{\rho\lambda}^3
  +\gamma^\rho\gamma^\nu{p\!\!\slash}_2\gamma^\mu{p\!\!\slash}_2 D_\rho^3
  +\gamma^\rho\gamma^\nu{p\!\!\slash}_2\gamma^\mu{p\!\!\slash}_1 D_\rho^3\big\}
 u(p_t)\epsilon_\nu^*(p_2)\epsilon_\mu^*(p_1)
\end{eqnarray}
with $T_1=T_{nm}^bT_{ml}^a$ ($n,l,a,b$ are respectively the color indices of the top, charm quark 
and the two gluons), $p_1$ and $p_2$ denoting the momenta of the two gluons,
and the four-piont loop functions'  dependence given by 
\begin{eqnarray}
D^1 &=&D^1(p_2, p_1 , -p_t, m_t, m_t, m_{\pi_t^0})\\
D^2 &=&D^2(p_2, p_1 , -p_t, m_t, m_t, m_{h_t^0})\\
D^3 &=&D^3(p_2, p_1 , -p_t, m_b, m_b, m_{\pi_t^+})
\end{eqnarray}
The amplitudes of the top-quark triangle loop digrams in Fig.2(e) are given by  
\begin{eqnarray}
M_1&=&\frac{1}{2}ag_s^2T_2\frac{i}{16\pi^2}\frac{i}{(p_1+p_2)^2-m_{\pi_t^0}^2}
  4m_t \bar{u}_c P_R\varepsilon^{\rho\lambda\nu\mu}p_{2\rho}
  p_{1\lambda}C_0\bar{u}_t\epsilon_\nu^*(p_2)\epsilon_\mu^*(p_1)\\
M_2&=&-\frac{1}{2}ag_s^2T_2\frac{i}{16\pi^2}\frac{i}{(p_1+p_2)^2-m_{\pi_t^0}^2}
  \bar{u}_cP_L m_t\big\{-4g_{\mu\nu}B_0  -16p_{1\nu}C_\mu
  + 16C_{\mu\nu} \nonumber \\
&& + 8p_{1}^\alpha C_\alpha g_{\mu\nu}
  -4p_1^2 g_{\mu\nu}C_0 -4p_1\cdot p_2 g_{\mu\nu}C_0
  +4p_{1\nu}p_{2\mu}C_0 \big\}\bar{u}_t\epsilon_\nu^*(p_2)\epsilon_\mu^*(p_1)
\end{eqnarray}
with $T_2=T_{nm}^bT_{ml}^a$ and the loop functions' dependence given by
\begin{eqnarray}
B =B(p_2^2 ,m_t^2,m_t^2), ~C =C(-p_1,-p_2,m_t,m_t,m_t) .
\end{eqnarray}
In the above expressions the loop functions $B$, $C$ and $D$ with Lorentz indices 
can be expanded into scalar functions \cite{Hooft},  which can be calculated
by using LoopTools \cite{Hahn}.


\begin{thebibliography}{99}
\bibitem{top-review} See, e.g.,
               D. Chakraborty, J. Konigsberg, D. Rainwater,  
                            Ann. Rev. Nucl. Part. Sci. {\bf 53}, 301  (2003);
               E.~H.~Simmons, hep-ph/0211335;
               C.-P. Yuan,  hep-ph/0203088;
               S. Willenbrock, hep-ph/0211067;
               M. Beneke {\it et al.}, hep-ph/0003033;
               C. T. Hill and S. J. Parke, \PRD49, 4454 (1994);
               K. Whisnant, et al.,  \PRD56, 467 (1997);
               K. Hikasa, et al., \PRD58, 114003 (1998).
               %%CITATION = HEP-PH 0303092;%%
               %%CITATION = HEP-PH 0211335;%%
               %%CITATION = HEP-PH 0203088;%%
               %%CITATION = HEP-PH 0211067;%%
               %%CITATION = HEP-PH 0003033;%%
               %%CITATION = PHRVA,D49,4454;%%
               %%CITATION = PHRVA,D56,467;%%
               %%CITATION = PHRVA,D58,114003;%%

\bibitem{tcvh-sm} For the FCNC top quark decays in the SM, see,
                  G.~Eilam, J.~L.~Hewett and A.~Soni, \PRD44, 1473 (1991);
                  B.~Mele, S.~Petrarca and A.~Soddu, \PLB435, 401 (1998);
                  A.~Cordero-Cid, {\it et al.}, \PRD73, 094005 (2006);
                  G. Eilam, M. Frank and I. Turan, \PRD73, 053011 (2006).
                  %%CITATION = HEP-PH 0411188;%%
                  %%CITATION = PHRVA,D44,1473;%%
                  %%CITATION = PHLTA,B435,401;%%
                  %%CITATION = PHRVA,D73,053011;%%

\bibitem{top-fcnc-review} For recent reviews, see, e.g.,
                        F.~Larios, R.~Martinez, M.~A.~Perez,
                        Int.\ J.\ Mod.\ Phys.\ A {\bf 21}, 3473 (2006);
                        J. M. Yang, Annals Phys. {\bf 316}, 529 (2005).
                        %%CITATION = HEP-PH 0605003;%%
                        %%CITATION = HEP-PH 0409351;%%

\bibitem{tcv-pptc-mssm} For the latest results of MSSM predictions for FCNC top decays and 
                        productions at LHC, see,  
      J. Cao, {\it et. al.}, \PRD75, 075021 (2007); \PRD74, 031701 (2006).
       %%CITATION = PHRVA,D75,075021;%%
       %%CITATION = PHRVA,D74,031701;%%
   
\bibitem{tcv-mssm}  For earlier studies on FCNC top decays in the MSSM, see,
    C.~S.~Li, R.~J.~Oakes and J.~M.~Yang, \PRD49, 293 (1994);
    G.~Couture, C.~Hamzaoui and H.~Konig, \PRD52, 1713 (1995);
    J.~L.~Lopez, D.~V.~Nanopoulos and R.~Rangarajan, \PRD56, 3100  (1997);
    G.~M.~de Divitiis, R.~Petronzio and L.~Silvestrini, \NPB504, 45 (1997);
    J.~M.~Yang, B.-L.~Young and X.~Zhang, \PRD58, 055001 (1998);
    C.~S.~Li, L.~L.~Yang and L.~G.~Jin, \PLB599, 92 (2004);
    M.~Frank and I.~Turan, \PRD74, 073014 (2006);
    %%
    J.~M.~Yang and C.~S.~Li, \PRD49, 3412 (1994);
    J.~Guasch and J.~Sola, \NPB562, 3 (1999);
    G. Eilam, {\it et al.}, \PLB510, 227 (2001).
    %%
    J.L. Diaz-Cruz, H.-J. He, C.-P. Yuan \PLB179,530 (2002);
    D. Delepine and S. Khalil, \PLB599, 62 (2004).
                    %%CITATION = PHRVA,D49,293;%%
                    %%CITATION = PHRVA,D52,1713;%%
                    %%CITATION = PHRVA,D56,3100;%%
                    %%CITATION = NUPHA,B504,45;%%
                    %%CITATION = PHRVA,D58,055001;%%
                    %%CITATION = PHLTA,B599,92;%%
                    %%CITATION = PHRVA,D74,073014;%%
                    %%CITATION = PHRVA,D49,3412;%%
                    %%CITATION = NUPHA,B562,3;%%
                    %%CITATION = PHLTA,B510,227;%%
                    %%CITATION = PHLTA,B179, 530;%%
                    %%CITATION = PHLTA,B599, 62;%%

\bibitem{pptc-mssm} Other works on top FCNC productions in the MSSM:
      J. Cao, Z. Xiong, J.M.Yang, \NPB651, 87 (2003);
      J.~J.~Liu, C.~S.~Li, L.~L.~Yang and L.~G.~Jin, \NPB705, 3 (2005);
      G.~Eilam, M.~Frank and I.~Turan, \PRD74, 035012 (2006);
      J. Guasch, {\it et al.}, Nucl. Phys. Proc. Suppl. 157, 152 (2006);
      D. Lopez-Val, J. Guasch, J. Sola, arXiv:0710.0587 
                    %%CITATION = PHRVA,D74,035012;%%
                    %%CITATION = NUPHA,B705,3;%%
                    %%CITATION = HEP-PH 0601218;%%
                    %%CITATION = NUPHA,B651,87;%%    
                    %%CITATION = ARXIV:0710.0587;%%

\bibitem{tcv-TC2}  For FCNC top quark decays in TC2 theory, see,
                   X.~L. Wang  {\it et al.}, \PRD50, 5781 (1994);
                   C.~Yue, {\it et al.}, \PRD64, 095004 (2001);
                   G.~Lu, F.~Yin, X.~Wang and L.~Wan, \PRD68, 015002 (2003).
                  %%CITATION = PHRVA,D50,5781;%%
                  %%CITATION = PHRVA,D64,095004;%%
                  %%CITATION = PHRVA,D68,015002;%%

\bibitem{pptc-TC2}  For exotic top production processes in TC2 models, see,
   H. J. He and C. P. Yuan, \PRL83, 28(1999);
   G. Burdman, \PRL83,2888(1999);
   J. Cao, Z. Xiong, J. M. Yang, \PRD67, 071701 (2003);
   C.~Yue, {\it et al.}, \PLB 496, 93 (2000);
   J. Cao, {\it et al.}, \PRD70, 114035 (2004);
   F. Larios and F. Penunuri, \JPG30, 895(2004);
   J. Cao, {\it et al.} \EPJC41, 381 (2005); \PRD76, 014004 (2007);
   G. Liu and H. Zhang, arXiv:0708.1553. 
                    %%CITATION = PRLTA,83,28;%%
                    %%CITATION = PRLTA,83,2888%%
                    %%CITATION = PHRVA,D67,071701;%%
                    %%CITATION = PHLTA,B496,301;%%
                    %%CITATION = PHRVA,D70,114035;%%
                    %%CITATION = HEP-PH 0311056;%%
                    %%CITATION = HEP-PH 0311166;%%
		    %%CITATION = PHRVA,D76,014004;%%
                    %%CITATION = ARXIV:0708.1553;%%

\bibitem{TC2-model}  C. T. Hill, \PLB345, 483 (1995);
              K. Lane and E. Eichten, \PLB352, 382 (1995);
              K. Lane and E. Eichten, \PLB433, 96 (1998);
              W.~A.~Bardeen, C.~T.~Hill, M.~Lindner, \PRD41, 1647 (1990);
              G. Cvetic, Rev. Mod. Phys. {\bf 71}, 513 (1999).
              %%CITATION = PHLTA,B345,483;%%
              %%CITATION = PHLTA,B352,382;%%
              %%CITATION = PHLTA,B433,96;%%

\bibitem{top-mass} Tevatron Electroweak Working Group
                  (for the CDF and D0 Collaborations),
                  hep-ex/0703034.
                  %%CITATION = HEP-EX 0703034;%%

\bibitem{kuang} C.~T.~Hill, X.~Zhang, \PRD51, 3563 (1995);
                C.~Yue, Y.~P.~Kuang, X.~Wang, W.~Li, \PRD62, 055005 (2000).
                %%CITATION = PHRVA,D51,3563;%%
                %%CITATION = PHRVA,D62,055005;%%

\bibitem{Hooft} G.~'t Hooft and M.~J.~G.~Veltman, \NPB153, 365 (1979).
                %%CITATION = NUPHA,B153,365;%%

\bibitem{Hahn} T.~Hahn, M.~Perez-Victoria, Comput.\ Phys.\ Commun.\ {\bf 118}, 153 (1999);
               T.~Hahn,  Nucl.\ Phys.\ Proc.\ Suppl.\ {\bf 135}, 333 (2004).
               %%CITATION = HEP-PH 0406288;%%
               %%CITATION = HEP-PH 9807565;%%
\end{thebibliography}
\end{document}